\documentstyle[12pt]{article}

\newcommand{\mysection}[1]{\setcounter{equation}{0}\section{#1}}

\newcommand{\nc}{\newcommand}
\nc{\beq}{\begin{equation}} \nc{\eeq}{\end{equation}}
\nc{\beqa}{\begin{eqnarray}} \nc{\eeqa}{\end{eqnarray}}
\nc{\lsim}{\begin{array}{c}\,\sim\vspace{-21pt}\\< \end{array}}
\nc{\gsim}{\begin{array}{c}\sim\vspace{-21pt}\\> \end{array}}

\begin{document}

\begin{titlepage}
\begin{center}
{\hbox to\hsize{July 1994 \hfill JHU-TIPAC-940013}}
{\hbox to\hsize{hep-ph/9407185 \hfill MIT-CTP-2343}}

\bigskip

\bigskip

{\Large \bf Low-Energy K\"ahler Potentials In Supersymmetric
Gauge Theories
With (Almost) Flat Directions} \\

\bigskip

\bigskip

{\bf Erich Poppitz\footnotemark[1]$^,$\footnotemark[2]}\\

\smallskip

{\small \it Department of Physics and Astronomy

The Johns Hopkins University

Baltimore, MD 21218, USA }

\smallskip
{\small and}
\bigskip

{\bf Lisa Randall}\footnotemark[3]\\

\smallskip

{ \small \it Center for Theoretical Physics

Laboratory for Nuclear Science and Department of Physics

Massachusetts Institute of Technology

Cambridge, MA 02139, USA }

 \bigskip

\vspace{2cm}

{\bf Abstract}\\[-0.05in]
\end{center}
We derive the supersymmetric low-energy effective theory of
the $D$-flat directions of a supersymmetric gauge theory.
The K\"ahler potential of Affleck, Dine and Seiberg is derived
by applying
holomorphic constraints which manifestly maintain supersymmetry.
 We also present a simple procedure for calculating all derivatives
of the
K\"ahler potential at points on the flat direction manifold.
Together with knowledge of the superpotential, these are sufficient
for a complete
determination of the spectrum and the interactions of the light
degrees of freedom. We illustrate the method  on the example of
a chiral abelian model, and comment on its application  to more
complicated calculable models with dynamical supersymmetry
breaking.

\bigskip

\footnotetext[1]{Supported in part by
NSF grants PHY90-96198 and PHY89-04035, and by
the Texas National Research Laboratory Commission
under grant RGFY93-292.}
\footnotetext[2]{After August 1, 1994:
{\it The Enrico Fermi Institute,
University of Chicago, 5640 Ellis Avenue, Chicago, IL 60637.}}
\footnotetext[3]{NSF Young Investigator Award,
Alfred P.~Sloan
Foundation Fellowship, DOE Outstanding Junior
Investigator Award. Supported in part
by DOE contract DE-AC02-76ER03069, by NSF grant PHY89-04035,
and by
the Texas National Research Laboratory Commission
under grant RGFY92C6.}

\end{titlepage}

\renewcommand{\thepage}{\arabic{page}}
\setcounter{page}{1}

\baselineskip=18pt

\mysection{Introduction}

When considering Lagrangians with many different mass scales,
it is often useful to integrate out the heavy degrees of freedom
and derive an effective low-energy theory of the light fields. In
supersymmetric theories, it is useful to follow this procedure
while manifestly preserving supersymmetry.

Supersymmetric gauge theories often exhibit directions in the
scalar field space, along which the scalar potential
identically vanishes, the so-called ``flat'' directions
\cite{ADS1}. When studying the low-energy dynamics
of the theory at a given point on the flat direction manifold
it is useful to consider an effective Lagrangian where the fields
are constrained to these  flat  directions \cite{ADS2}.
However, the vanishing  of the $D$ term is a nonholomorphic constraint
 which  generally cannot be solved in terms of chiral
superfields. It is therefore not clear a priori how to construct
an effective theory of the flat directions which manifestly
maintains supersymmetry.

The method for finding the K\"ahler potential of the low-energy sigma
model used by Affleck, Dine and Seiberg (ADS) \cite{ADS2} is based on
the procedure of
 using the flat direction equations to project the full theory
K\"ahler potential without gauge fields onto the light gauge invariant
chiral superfields.
For simple gauge groups and matter representations, such as  SU(2)
one-flavor SQCD, it leads to unambiguous results for the
K\"ahler potential.  However,  even in the case of the simplest
model with dynamical supersymmetry breaking, the SU(3)$\times$SU(2)
model \cite{ADS2}, the K\"ahler potential is determined by the solution
of a cubic equation \cite{ADS2}. The choice of the correct solution
can only be made by examining the positivity of the K\"ahler metric
for each of  the roots of this equation \cite{us}.

In this paper, we address both of these ambiguities. We derive the ADS
effective Lagrangian by applying a {\it holomorphic} constraint and
explicitly integrating out the heavy vector fields. We show how
this procedure is equivalent, via a {\it nonholomorphic} field
redefinition,
 to the ADS procedure.

We also show how to compute all derivatives of the K\"ahler potential at
the flat direction without
solving the complicated equations for the vanishing of the $D$ terms.
Moreover, our procedure
 yields a
manifestly positive definite K\"ahler  metric, as we show in  Sect. 2. It
 is very general and can be applied to
calculable models of dynamical supersymmetry breaking,  with virtually
unknown
ground state  properties.

\mysection{Gauge Invariant Description of the Light Degrees of Freedom}

In this section we derive the gauge invariant theory of the light
degrees of freedom along a flat direction with completely broken
gauge symmetry.
It can be generalized to the case where there is some unbroken
nonabelian
gauge group, so long as we are interested in the effective theory
at scales below the scale where all particles carrying charge under
the unbroken group acquire a dynamical mass.
The matter part of the classical Lagrangian of a general
supersymmetric gauge theory with gauge group $G$
(of dimension $N - n$) and no  classical superpotential
 has the form of a D-term:
\beq
\label{su2qcd}
L_D ~ = ~ Q^{\dagger} ~ e^{V^a T^a} ~ Q ,
\eeq
where $Q_i$ are  chiral matter superfields. Hereafter $i$ runs
over both the gauge index and the different representations,
$i = 1,...,N$, where $N$ is the
 number of chiral matter superfields. $V^a$,
$a = 1,...,N-n,$ denote the vector superfields, corresponding to
the various factors in $G$. Under a supergauge transformation the
matter and gauge superfields transform as \cite{WB}:
\beqa
\label{transformations}
Q~&\rightarrow &~e^{- i \Omega^a T^a}~Q  \\
 e^{V^a T^a} ~&\rightarrow &~e^{-i \Omega^{a \dagger} T^a}~e^{V^b T^b}~
e^{i\Omega^a T^a}~\nonumber,
\eeqa
where the parameters of the transformation $\Omega^a$ are chiral
superfields. The scalar potential of this theory has classically
flat directions
along which it identically vanishes. They are given by the solutions of
\beq
\label{flatdirections}
Q^{\dagger i}~T^{a ~j}_i~Q_j~ =~0 ,
\eeq
where $Q_i$ now means the scalar components of the corresponding chiral
 superfields
and a sum over the different representations is again implicit. If we
expand the theory around a solution of (\ref{flatdirections}) sufficiently
far from the
origin  the theory is   weakly coupled and can be analyzed
perturbatively \cite{ADS1}. We consider the theory  in the vicinity of a
such  solution of   (\ref{flatdirections}), which completely breaks the
gauge symmetry \cite{ADS2}
(up to possible abelian factors).
The number of broken generators of the gauge group is $N - n$.
Then $N - n$ of the $N$ chiral superfields $Q_i$ are massive
and $n$ chiral superfield are  massless (in the absence of superpotential).

Below the scale of the gauge boson  masses the heavy gauge fields
and their superpartners can be integrated out.
As in ref.\cite{ADS2} we assume that the theory of the light degrees
of freedom can be given a gauge invariant description, where the light
supermultiplets are represented by a set of gauge invariant chiral
superfields $X^A$,  $A = 1,...,n$, which are independent polynomials
in the matter fields $Q_i$:
\beq
\label{lightfield}
X^A ~= ~ X^A (Q).
\eeq
 The dynamics of the theory below the scale of the gauge boson masses
is described by a supersymmetric K\"ahler sigma model with coordinates
spanned by the light chiral superfields (\ref{lightfield})
\cite{ADS2}. The derivation of the K\"ahler potential of this sigma
model is the focus of this letter.

The derivation is nontrivial, because if one follows the procedure of
ref.\cite{ADS2} and
applies the nonholomorphic constraint of eq.\ref{flatdirections}  it
is not clear why the resulting Lagrangian should be supersymmetric.
Explicitly, for particular models one observes that
 the flat directions equations
cannot be promoted to chiral superfields.
Furthermore, because the constraint equations are real, there is
an insufficient number of constraining equations to determine the
light chiral  fields (see discussion below).

Another difficulty with the ADS procedure is more a matter of
practice. To find the K\"ahler potential of the effective theory,
one needs to solve for it along the flat directions, which
can be very complicated.
Even for the simplest model of dynamical supersymmetry breaking
based on ${\rm SU}(3) \times {\rm SU}(2)$, one needs to solve
a cubic polynomial equation and only the correct root gives
a positive definite kinetic energy for the light degrees of freedom.

In the rest of this section, we  will show how to derive the ADS
potential. We will use
holomorphic constraints when separating the light from the heavy
degrees of freedom.  We will show using our procedure how
one can compute derivatives of the K\"ahler potential
(which are all that are required for finding the spectrum
and interactions) simply, without explicitly solving for
the full form of the low energy K\"ahler potential.

In order to separate the light and heavy degrees of freedom in a gauge
invariant way, it is necessary to make a field redefinition
\beqa
\label{fieldredefinition}
Q~&\rightarrow &~e^{- i G^a T^a}~Q(X)   \\
e^{V^a T^a}~&\rightarrow &  ~e^{- i G^{c \dagger} T^c}~e^{{\cal{V}}^a T^a}~
e^{i G^b T^b} ~,\nonumber
\eeqa
where $G^a$ are Goldstone chiral superfields, transforming as
\beq
\label{goldstonetransform}
e^{- i G^a T^a} ~\rightarrow ~ e^{- i \Omega^a T^a}~ e^{- i G^b T^b}
\eeq
under supersymmetric gauge transformations. The  vector superfield
${\cal{V}}^a$ is  gauge invariant, as follows
from (\ref{transformations}), (\ref{fieldredefinition}), and
(\ref{goldstonetransform}).
On-shell it describes
a massive vector supermultiplet \cite{WB}.

We are interested in making the field redefinition (\ref{fieldredefinition})
locally,
 around a   point $Q_i = v_i$
on the flat direction
manifold (``moduli space'')  where the gauge symmetry
is completely broken. Hence the $N-n$ vectors ($N$-dimensional)
$T^{a ~j}_i v_j$ are all nonvanishing and linearly independent.
We assume that one can find $X^A (Q)$ such that all derivatives
$X^A_i (Q)  \equiv \partial X^A (Q)/\partial Q_i$
are nonvanishing at this point. The functional independence of
$X^A (Q)$ then assures that the $n$ vectors ($N$-dimensional)
$X^A_i (v)$ are linearly independent. Gauge invariance of $X^A (Q)$ implies
that $X^A_j (v) T^{a ~j}_i v_j = 0$, for any $A, a$.
Taken together with the linear independence, this implies that
 the $N$ vectors
$X^{A *}_i (v^*) \equiv
\partial X^{A *} (Q^*)/\partial Q^{* i}\vert_{Q^*=v^*}$ and
  $T^{a ~j}_i v_j$ form a complete basis in the complex space
spanned by $Q_i$.
The gauge invariant fields $X^A$ have the expansion around $Q_i = v_i$
\beq
\label{Xexpanded}
X^A = X^A(v_i) + x^A ~.
\eeq
Then  the  expansion of $Q(X)$ around  points on the flat direction manifold
 is
 \beq
\label{Qexpanded}
Q_i (X) ~=~ v_i + q_i (x)~.
 \eeq
The most general form of
$q_i(x)$ is
\beq
\label{qiofx}
q_i (x)~=~ X^{A *}_i (v) \lambda_A (x)~+~T^{a ~j}_i v_j \chi_a (x)~.
\eeq

An important application of this formalism is to theories which possess
flat directions only in a certain limit, the so-called almost flat
directions.
Examples
of such models are massive supersymmetric QCD (with masses much smaller
than the strong coupling scale of the theory) \cite{ADS1}, the
SU(2)$\times$SU(3) \cite{ADS2}, and
the calculable SU(5) \cite{ADS3} models of dynamical supersymmetry breaking.
For our purposes, the common
property of these models is that the superpotential can be considered as a
perturbation so long as the scale of the vacuum expectation values
along the flat directions is larger than the strong coupling scale of the
theory.
The superpotential  is a gauge invariant
 holomorphic function \cite{SV} of the chiral superfields $Q_i$,
$W = W(X(Q))$ .
After the field redefinition (\ref{fieldredefinition}), by gauge invariance
of $W$, the resulting superpotential is independent of the Goldstone
superfields:
\beq
\label{superpotential}
 W~=~W(X(Q(X)))~=~W(X)~.
\eeq
Here we required that $Q(X)$ obey
\beq
\label{xofqofx}
X(Q(X))~=~X ~.
\eeq
Notice that requiring (\ref{xofqofx})
allows a nonholomorphic ${\cal{Q}}(X^{\dagger},X)$ of a special form
\beq
\label{specialq}
{\cal{Q}} (X^{\dagger}, X)~=~ e^{{\cal{B}}^a (X^{\dagger}, X) T^a}~Q (X)~,
\eeq
with ${\cal{B}}^a (X^{\dagger}, X)$ an arbitrary complex function. Since
$X(Q)$ is invariant under the complex extension of the gauge group,
the nonholomorphic factor disappears from (\ref{xofqofx}).

With the most general form of $Q_i (X)$ given in terms of the $N$ functions
  $q_i (x)$ (\ref{qiofx}),
the $n$ holomorphic functions $X(Q)$ can only be inverted after imposing
$N - n$
holomorphic constraints. The field redefinition (\ref{fieldredefinition})
amounts to introducing new coordinates on the space of the chiral
superfields
spanned by $Q_i$:
\beq
\label{coordinate}
\{ Q_i \} ~\rightarrow~ \{ X^A (Q), S^a (Q) \}~,
\eeq
the only condition on $S^a (Q)$ being the nondegeneracy of the mapping
(\ref{coordinate}):
\beq
{\rm det}\left( {\partial X^A (Q)\over \partial Q_i},
{\partial S^a (Q)\over \partial Q_i}\right) \ne 0~.
\eeq
 In the vicinity of $Q_i = v_i$
the additional $N - n$ coordinates $S^a$ are
\beq
\label{SA}
S^a~=~ v^{* j} T_j^{a ~i} Q_i~,
\eeq
since
${\rm det}( X^A_i (v), v^{* j} T_j^{a ~i}) \ne 0$.
Now the holomorphic functions $X(Q)$, obeying (\ref{xofqofx}) can be
inverted by imposing $N-n$ holomorphic constraints. We choose the constraints
\beq
\label{SA1}
S^a~=~ v^{* j} T_j^{a ~i} Q_i~=~0,
\eeq
since then the fields $Q(X)$ (\ref{qexpanded}) are flat to linear order,
$Q^* T^a Q = 0 + O(x^2)$ (this follows from  gauge invariance
of $X^A(Q)$).

Notice that this equation can be interpreted as the fact that the
projection of the light field on the Goldstone boson direction vanishes.
Then the fields
$Q_i$ (\ref{Qexpanded}), (\ref{qiofx}) obeying the constraint (\ref{SA1})
have the expansion
\beq
\label{qexpanded}
 Q_i (X)~=~ v_i~ + ~X^{A *}_i \lambda_A (x) .
\eeq
 The $n$ functions $\lambda_A (x)$ are determined by inverting (\ref{xofqofx})
\beq
\label{definelambda}
x^A~=~X^A (v_i + X^{B * }_i \lambda_B (x)) ~-~X^A (v_i)~,
\eeq
 in terms of a Taylor series in $x^A$:
\beq
\label{lambdaexp}
\lambda_A (x)~ =~ (X^C_i X^{* D}_i)^{-1}_{~ A B} ~x^B ~+~O(x^2) .
\eeq

Expressing the Lagrangian (\ref{su2qcd}) in terms of the new fields
$Q(X)$ and $\cal{V}$,
and expanding in powers of the heavy field $\cal{V}$, we obtain
\beqa
\label{expandedLagrangian}
L_D~&=&~Q(X)^{\dagger}~ e^{{\cal{V}}^a T^a}~Q(X)    \\
&=&~Q(X)^{\dagger}~Q(X) ~+
{}~{\cal{V}}^a~ Q(X)^{\dagger}~T^a~Q(X)~ +~...~~,\nonumber
\eeqa
where dots denote higher powers of the massive vector
superfield\footnote{The gauge invariant field $\cal{V}$, defined by
(\ref{fieldredefinition}), includes the Goldstone
fields and their superpartners and therefore there are
infinitely many terms in the expansion of $e^{\cal{V}}$.}.
Below the scale of the mass of the gauge fields, the heavy vector
multiplet can be integrated out. Since we are only interested in
the leading term of the low-energy expansion, we can neglect the
kinetic term of the gauge field.
The
zero-momentum tree graphs are easily computed by perturbatively solving
the equation of motion for $\cal{V}$ that follow from
(\ref{expandedLagrangian}).  Then the low-energy K\"ahler potential
 $K(X^{\dagger}, X)$ is given by
(\ref{expandedLagrangian}) with the field ${\cal{V}}_c$ substituted
by the solution to its classical equation of motion, i.e.
\beq
\label{lowenergykahler}
K~=~Q^{\dagger} (X)~e^{\cal{V}}~Q (X)\big\vert_{{\cal{V}} =
{\cal{V}}_c}~,
\eeq
with ${\cal{V}}_c$ determined by
\beq
\label{vc}
  {d K\over d {\cal{V}}}\bigg\vert_{ {\cal{V}} = {\cal{V}}_c} ~=~ 0 ~.
\eeq

If the functions $Q(X)$ obey the flat direction equations
(\ref{flatdirections})  there are no additional tree-level contributions
to the low-energy K\"ahler potential, as follows from
(\ref{expandedLagrangian})\footnote{In the  one-flavor SU(2) SQCD finding
such a redefinition is simple.
Let $Q$ and $\bar{Q}$ be the two doublets of chiral matter transforming
in the
$2$ and $2^*$ representations.
For dimensional reasons $Q(X)~=~\sqrt{X} \eta $ and
$\bar{Q}(X)~=~\sqrt{X} \xi $,
with $\eta $ and $\xi $ being two constant  spinors with unit norm,
obeying $\eta^{\dagger}~T^a~\eta ~-~ \xi ~T^a~ \xi^{\dagger}~=~0 $.
Then the K\"ahler potential  $K(X^{\dagger}, X)$ becomes
$2 \sqrt{X^{\dagger}~X}$, which coincides with the potential derived
in \cite{ADS2}.}.
In general, however, the holomorphic functions (\ref{qexpanded}),
(\ref{definelambda}) do not obey the flat directions equations.
This is the case, e.g. in the SU(3)$\times$SU(2) model with dynamical
supersymmetry breaking \cite{us} and in the simple abelian chiral
model considered in the next section.
 Then  there are  zero-momentum tree-level graphs due to the heavy
vector
supermultiplet, which contribute to the low-energy K\"ahler potential.

Since the holomorphic functions $Q(X)$ are flat up to quadratic
 order in the expansion around the flat direction, the leading contribution
of the tree graphs is fourth order in $x^A$. So when computing
derivatives of fourth order or higher one needs to incorporate
the contribution to the potential from the nonzero vector field.
However, to compute the second and third derivatives, it is sufficient
to invert $Q$ in terms of $X$.
The K\"ahler metric at the minimum does not receive tree-level corrections
from the heavy fields and is  given by the manifestly positive definite
expression
\beq
\label{kahlermetric}
K_{A^* B}~=~~ (X^{C}_i X^{* D}_i)^{-1}_{~A  B}~ ~ .
\eeq
Similarly, using (\ref{qexpanded}), (\ref{definelambda}) and
(\ref{lowenergykahler}), one can derive expressions involving three
derivatives.
However, because the $Q$ fields we define are not $D$-flat,
to calculate fourth or higher order  derivatives of the K\"ahler
potential at the minimum, one needs to incorporate
the additional contribution from integrating out the vector field.

However, there does exist a  nonholomorphic redefinition
of the fields such that  $\tilde Q_i (X^{\dagger}, X)$
obey  the (nonholomorphic) flat direction equations
(\ref{flatdirections}). The K\"ahler potential (\ref{lowenergykahler})
 can  be written as
\beq
\label{k1}
K~=~Q^{\dagger} (X)~e^{{\cal{V}}_c /2}~e^{{\cal{V}}^{\prime}}~
e^{{\cal{V}}_c /2}~Q (X)~\bigg\vert_{{\cal{V}}^{\prime} = 0}~
\eeq
 Defining the new fields
\beq
\label{nonholomorphic}
\tilde{Q} (X^{\dagger}, X)~\equiv~ e^{{\cal{V}}_c (X^{\dagger}, X) /2}
{}~Q (X) ~,
\eeq
the K\"ahler potential (\ref{k1}) becomes
\beq
\label{k2}
\tilde{K} ~=~\tilde{Q}^{\dagger} (X^{\dagger}, X)~e^{{\cal{V}}^{\prime}}~
\tilde{Q} (X^{\dagger}, X)\bigg\vert_{{\cal{V}}^{\prime} = 0}~.
\eeq
Now, from (\ref{k2}) it follows that
\beq
\label{flat}
{d \tilde{K} \over d {\cal{V}}^{\prime a}}\bigg\vert_{{\cal{V}}^{\prime}
= 0}~=
 ~\tilde{Q}^{\dagger} (X^{\dagger}, X)~T^a ~\tilde{Q} (X^{\dagger}, X)~=
{}~{d K\over d {\cal{V}}}\bigg\vert_{ {\cal{V}} = {\cal{V}}_c} ~=~ 0 ~,
\eeq
where the second equality follows from the redefinition
(\ref{nonholomorphic})
 and the last holds by virtue of (\ref{vc}). Equation (\ref{flat}) shows
that the nonholomorphic  functions (\ref{nonholomorphic})
$\tilde{Q} (X^{\dagger}, X)$
obey the flat direction equations.
These fields correspond to those of Affleck, Dine, and Seiberg,
\cite{ADS1},
and differ from the fields $Q$ (\ref{qexpanded}) at quadratic
order in $x^A$.

Note that the $D$-flat conditions only give
$N-n$   real constraints whereas we need
$N-n$  complex  ones,
hence there is an insufficient number of constraints.
In the ADS construction  the $N-n$ additional
real constraints correspond to fixing the gauge.

The important fact is that even though the fields $\tilde{Q}$
are not holomorphic,
the superpotential constructed from these
fields is nonetheless manifestly supersymmetric.
This is because the  low-energy superpotential is
still a holomorphic function of the light superfields, since
by gauge invariance, the nonholomorphic factor
$e^{{\cal{V}}_c (X^{\dagger}, X) /2}$ in the redefinition
drops out of the superpotential. This justifies the ADS construction.

Therefore, when deriving the effective theory, one is
faced with several possibilities. One can apply a holomorphic
constraint to restrict oneself to the light degrees of freedom.
It is easy with this procedure to derive all derivatives of
the K\"ahler potential\footnote{Alternatively, one can
explicitly solve the holomorphic constraint equation, similar
to the procedure of ADS (as in the one-flavor SU(2) example above).}.
When applying this procedure to the
derivatives of fourth or higher order, one needs to include
the contribution from explicitly integrating out the heavy
vector fields by solving their equations of motion.

Alternatively, one can use nonholomorphic fields which
exactly satisfy the $D$ flat equations. This procedure is
justified by the fact that the nonholomorphic field contribution
will cancel out from gauge invariant superpotential terms. Here,
the classical vector field is zero, so there are no additional
contributions to the K\"ahler potential for the light fields.
The leading order derivative terms are the same as those
calculated with the previous procedure. The higher order
terms can be computed perturbatively as well; these
are the same as those derived from the Lagrangian with a holomorphic
constraint, so long as the classical vector field contribution
is incorporated.

We should also note that if there is some unbroken nonabelian gauge group,
 our procedure can be applied, as long as we are interested in the
effective theory below the dynamically generated mass scale of
the fields carrying gauge charge with respect to the
unbroken gauge group.
So long as we can find $X^A(Q)$ such that
$T_j^{\alpha ~i} X^A_i (v) = 0$ holds for the
unbroken generators $T^{\alpha}$,
it is easy to see that the fields (\ref{qexpanded})
obey the unbroken group $D$-flat equations and   are
decoupled from the (strongly interacting) gauge field of the
unbroken group.

\mysection{The Low-Energy K\"ahler Potential: an Abelian Example}

In this section we illustrate our method on a simple abelian chiral
model.
The advantage is that we can solve  the vector field equations of
motion
to all orders and explicitly demonstrate
the equivalence of our procedure with that of ADS via a nonholomorphic
field redefinition.

Consider a chiral U(1) supersymmetric gauge theory with three
chiral matter superfields: $S_1$, $S_2$ of charge 1, and  $T$,
of charge -2.
The Lagrangian of the model   is
\beq
\label{chiralu1}
L_D~=~S_1^{\dagger}~e^V~S_1~+~~S_2^{\dagger}~e^V~S_2~+~~T^{\dagger}~
e^{-2 V}~T~.
\eeq
Along the flat directions, given by
\beq
\label{abelianflatdirections}
S_1^{\dagger} S_1~+~S_2^{\dagger} S_2~-~ 2 T^{\dagger} T~=~0 ,
\eeq
the gauge symmetry is broken.
 There are two independent gauge invariant chiral superfields in
this theory
\beq
\label{abelianinvariants}
X_1~ = ~S_1^{~ 2}~T ~, ~  ~X_2~ = ~S_2^{~ 2}~T ~.
\eeq
Performing the field redefinition (\ref{fieldredefinition}),
with $Q_i(X)$ determined by (\ref{qexpanded}), (\ref{definelambda}),
the low-energy K\"ahler potential is given by (\ref{chiralu1}),
where the
corresponding functions $S_1 = S_1(X_1,X_2)$, etc., are substituted,
and $V$
denotes the heavy vector multiplet. Substituting the solution to the
equation of motion for
$V$,
\beq
\label{Vsaddle}
e^V~=~\left( 2 T^{\dagger} T\over S_1^{\dagger} S_1 +
S_2^{\dagger} S_2 \right)^{1/3}~,
\eeq
into (\ref{chiralu1}), the low-energy K\"ahler potential becomes
\beq
\label{kahler1}
K( X^{\dagger}, X ) ~=~
3~\left(~{\sqrt{(S_1^{~ 2} ~T)^{\dagger}~ S_1^{~ 2}~ T }~+~
\sqrt{ (S_2^{~ 2}~ T)^{\dagger}~ S_2^{~ 2}~ T } \over 2} ~\right)^{2/3} ~.
\eeq
Finally, recalling that $S_1(X_1,X_2)$, etc., obey (\ref{xofqofx}),
e.g.
$S_1^{~ 2} (X_1,X_2)~T (X_1,X_2) = X_1$, we find the expression for
the K\"ahler potential of the light degrees of freedom
\beq
\label{abeliankahlerADS}
K( X^{\dagger}, X ) ~=~
3~\left(~{\sqrt{X_1^{\dagger} X_1 }~+~\sqrt{ X_2^{\dagger} X_2  }
         \over 2} ~\right)^{2/3} ~.
\eeq
This coincides with the K\"ahler potential obtained by the method of
\cite{ADS2}.

Notice that the vector field contribution only affected fourth
and higher order derivatives, since its expansion in light fields
begins at second order, as follows from the fact that the
fields (\ref{qexpanded}) are $D$ flat to linear order. In more
complicated
examples, where one cannot explicitly solve the vector field
equation of motion to all orders, one can nonetheless perturbatively
derive
the vector field contribution. Explicitly, the scalar contribution
corresponds to integrating out the auxiliary complex scalar
  component (which vanishes in the Wess-Zumino gauge; it is
 denoted by $M + i N$ in ref.\cite{WB}) of the vector field.

\mysection{Conclusion}

In this letter we developed a procedure for finding the K\"ahler
potential of the
light degrees of freedom in supersymmetric theories, where the gauge
symmetry is completely broken along a flat direction of the $D$-term
scalar
potential. The resulting K\"ahler potential is determined as a power
series expansion around the given point of the flat direction.
The K\"ahler
metric is manifestly positive definite.

The method is quite general and can be applied to any calculable
model
of dynamical supersymmetry breaking.
It is satisfying that one can derive the low energy
theory without exactly solving the flat directions equations in terms
of the gauge invariant superfields, particularly
in the case of more complicated models. This might
prove useful when deriving the physics of specific models.
Of particular interest is the SU(5) model
with two generations \cite{ADS3}, since little is known about
its ground state or
broken
symmetries.

 \section{Acknowledgements}

We are grateful to Jon Bagger, Ken Intriligator and Joe Polchinski
for useful discussions, and the ITP at Santa Barbara for hospitality
during the workshop{\it Weak Interactions '94}.


\begin{thebibliography}{99}

\nc{\ib}[3]{ {\em ibid. }{\bf #1} (19#2) #3}
\nc{\np}[3]{ {\em Nucl.\ Phys. }{\bf #1} (19#2) #3}
\nc{\pl}[3]{ {\em Phys.\ Lett. }{\bf #1} (19#2) #3}
\nc{\pr}[3]{ {\em Phys.\ Rev. }{\bf #1} (19#2) #3}
\nc{\prep}[3]{ {\em Phys.\ Rep. }{\bf #1} (19#2) #3}
\nc{\prl}[3]{ {\em Phys.\ Rev.\ Lett. }{\bf #1} (19#2) #3}

\bibitem{ADS1}I. Affleck, M. Dine and N. Seiberg, \np{B241}{84}{493}.

\bibitem{ADS2}I. Affleck, M. Dine and N. Seiberg, \np{B256}{85}{557}.

\bibitem{us}J. Bagger, E. Poppitz and L. Randall, {\it The $R$
Axion From Dynamical Supersymmetry Breaking}, preprint
JHU-TIPAC-940007,
MIT-CTP-2309, NSF-ITP-94-48; {\it Nucl. Phys} {\bf B}, in press.

\bibitem{WB}J. Wess and J. Bagger, {\it Supersymmetry and
Supergravity},
(Princeton University Press, Princeton, NJ, 1992).

\bibitem{ADS3}I. Affleck, M. Dine and N. Seiberg, \prl{52}{84}{1677}.

\bibitem{SV}M.A. Shifman and A.I. Vainshtein, \np{B277}{86}{456},
\np{B359}{91}{571};

N. Seiberg, \pl{B318}{93}{469};

M. Dine and Y. Shirman, {\it Some Explorations in Holomorphy},
preprint SCIPP-94/11.

\end{thebibliography}
\end{document}